\documentclass[preprint2]{aastex}

\shorttitle{Population effects on MDFs from the RGB}
\shortauthors{Ordo\~{n}ez \& Sarajedini}

\begin{document}

\title{Population Effects on the Metallicity Distribution Function Derived 
From the Red Giant Branch}

\author{Antonio J. Ordo\~{n}ez\altaffilmark{1}}
\email{a.ordonez@ufl.edu}

\and

\author{Ata Sarajedini\altaffilmark{1}}
\email{ata@astro.ufl.edu}

\altaffiltext{1}{Department of Astronomy, University of Florida, 211
  Bryant Space Science Center, Gainesville, FL 32611, USA}

\begin{abstract}
We have tested the reliability of the red giant branch (RGB) as a metallicity
indicator accounting for observational errors as well as the
complexity of star formation histories (SFHs) and chemical evolution
histories observed in various stellar systems. We
generate model color-magnitude diagrams (CMDs) produced with a variety of
evolutionary histories and compare the resultant metallicity estimates from the 
colors and magnitudes of RGB stars
to the true input metallicities. We include realistic models for
photometric errors and completeness in our synthetic CMDs. As expected, 
for simple stellar populations dominated by old stars, the RGB
provides a very accurate estimate of the modular metallicity value for
a population. An error in the age of a system targeted for this type
of study may produce metallicity errors of a few tenths of a dex. The
size of this metallicity error depends linearly on the age error, and
we find this dependence to be stronger with more precise photometry. If the population has experienced any
significant star formation within the last $\sim$6 Gyr, the
metallicity estimates, [M/H], derived from the RGB may be in error by
up to
$\sim$0.5 dex. Perhaps the most important consideration for
this technique is an accurate, independent estimate of the average age for the target
stellar system, especially if it is probable that a significant
fraction of the population
formed less than $\sim$6 Gyr ago. 
\end{abstract}


\section{Introduction}
\label{sec:intro}
The red giant branch (RGB) phase in intermediate- to low-mass stellar evolution
corresponds to the phase just before the onset of helium
fusion. During the RGB phase, a star contains an electron
degenerate core surrounded by a hydrogen-burning shell fueled by the
CNO cycle \citep{chi98}. 

The color of the RGB on a
color-magnitude diagram (CMD) strongly depends on the metal abundance
of the stellar population (e.g. \citet{da90}). For this reason, 
the RGB has been widely used to obtain average metallicity estimates for a 
variety of stellar systems in the form of straightforward empirical calibrations 
between magnitude, color, and metallicity \citep{da90,sav00, str14}. However, the 
morphology and position of the RGB also depend 
somewhat on the age of the population, especially for stellar
populations younger than $\sim$10 Gyr. Thus, given an estimate for the age of a stellar population,
the metallicity distribution function (MDF) of the system can be calculated by comparing the
RGB with theoretical isochrones of the appropriate age \citep{sar05, mcc06} or fiducial RGB
sequences for discrete ages \citep{dur01}. 

Previously, \citet{sal05} had investigated biases in determining the
tip of the red giant branch (TRGB) based in the galaxies LMC, SMC,
and LGS3 using synthetic CMDs. That study investigated some population
effects on metallicities derived from the synthetic CMDs representing
these systems.
In this work, we provide a more thorough investigation of the degree to which star formation history can influence 
the results of these types MDF analyses mentioned above. To facilitate
this, we generate model CMDs with a variety of star
formation histories (SFHs) and chemical evolution histories and
calculate the MDFs by interpolating amongst a grid of isochrones. In
Section \ref{sec:meth}, we outline the methods used, detailing how the
synthetic CMDs were created, how error and completeness profiles were
applied, and how the interpolations were performed. Results are
presented in Section
\ref{sec:res}, and we conclude our findings in Section \ref{sec:con}.

\section{Methods}
\label{sec:meth}
\subsection{Synthetic CMDs}
\label{sub:scmd}
To generate the synthetic CMDs, we utilize
IAC-STAR\footnote{http://iac-star.iac.es} \citep{apa04}. This web browser-based 
software package
generates synthetic CMDs through bilogarithmic interpolation in age
and metallicity for a given set of stellar evolution
libraries. IAC-STAR allows the user to enter unique SFHs and
age-metallicity relations (AMRs) to synthesize arbitrarily complex stellar
populations. It does so by taking as input up to 20 nodes in time
with corresponding star formation rates (SFRs) and metallicities. The
code then performs interpolations between these nodes to achieve
satisfactory temporal resolution. 

The
stellar properties are calculated for each star and then converted to
absolute magnitudes in a range of different filter sets with a variety
of bolometric correction libraries. We utilize the stellar evolution
library of \citet{gir00} along with the bolometric correction library
from \citet{gir02}. We chose this combination after our own
exploration revealed that the RGBs produced from IAC-STAR with the
stellar evolution library of \citet{gir00} more closely matched the
corresponding isochrones than the other stellar evolution libraries
made available. We retrieved the isochrones to construct the
interpolation grid using the CMD v2.1 web
interface\footnote{http://stev.oapd.inaf.it/cgi-bin/cmd\_2.1} since
these utilize the same bolometric correction library used by IAC-STAR.
For simplicity, we use a distance modulus of zero, zero reddening, a
Kroupa IMF \citep{kro01}, and an age of 13.5 Gyr for the universe in all of our
models. We also set the mass-loss parameters for the RGB and AGB to
$\eta = 0.2$ since we focus on old (low-mass) populations. As we
are interested in magnitudes consistent with the brighter RGB
($-4\lesssim$M$_{\mathrm I}\lesssim0$) in the present study, 
we set the  limit for output from IAC-STAR to M$_{\mathrm V} = 2$ mag. 

\subsection{Photometric Errors and Completeness}
\label{sub:err}
In order to realistically model the observational characteristics of
CMDs, we follow the prescription of \citet{bar04}. In that work,
they tested the reliability of the TRGB
using synthetic CMDs where they modeled the photometric error and
completeness profiles using simple analytic functions. Photometric errors 
for both M$_{\mathrm{V}}$ and M$_{\mathrm{I}}$ are modeled with an
exponential function:
\begin{equation}\label{eq:err}
\sigma(\mathrm{M}) = \kappa e^{\tau \mathrm{M}}
\end{equation}
where $\kappa$ and $\tau$ are coefficients that describe the shape of
the error profile. The completeness at a given magnitude is modeled as:
\begin{equation}\label{eq:compl}
f(\mathrm{M}) = -\frac{2}{\pi} \arctan [\alpha (\mathrm{M} - \mathrm{M}_0)]
\end{equation}
where $\alpha$ is a shape parameter and M$_0$ is the magnitude at which completeness falls to 0\%. 
\begin{figure}
\epsscale{1.2}
\plotone{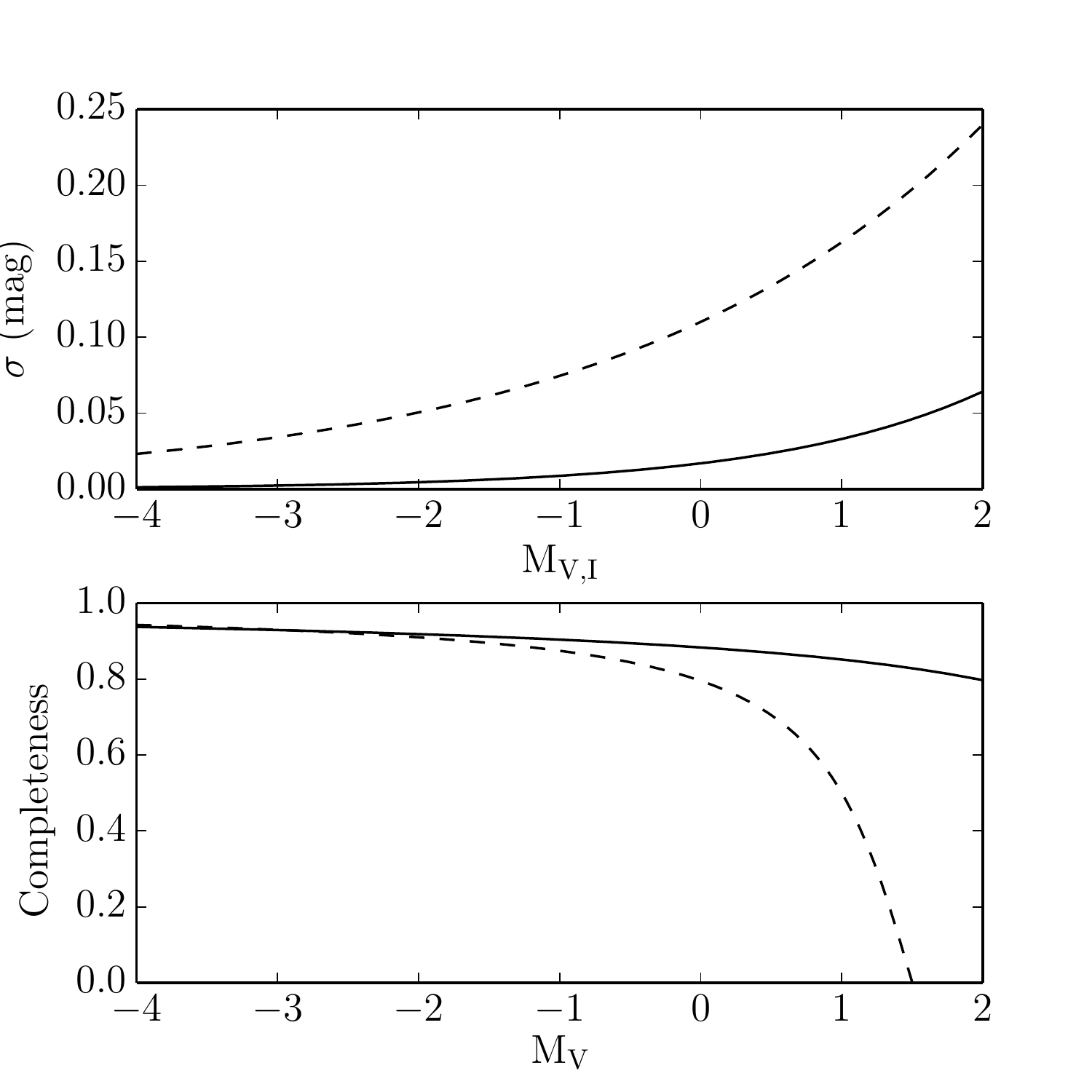}
\caption{\small{Both adopted error and completeness profiles. \emph{Top:}
  Photometric error as a function of absolute magnitude, both in V and
  I. The dashed line corresponds to Profile A, while the solid
  line corresponds to Profile B. \emph{Bottom:} Completeness as a
  function of M$_{\mathrm{V}}$. Symbols are as for the top panel.\label{fig:err}}}
\end{figure}

In an effort to sample two extremes in error and completeness, we
define two error and completeness profiles in the following
way. For the first case, we utilize the same error and completeness profiles from
\citet{bar04}, which will be denoted as observational profile A. The relevant
coefficients for this error profile are listed in Table
\ref{tbl:err}. This results in typical errors on the RGB of $\le$0.05
mag at absolute magnitudes brighter than M = -2 mag and a completeness of 50\% at
M$_{\mathrm{V}} = 1$ mag.  A typical CMD obtained using this
  profile contains $\sim$2200 stars.

The second case is
aimed at representing a less conservative error profile similar to
\emph{Hubble Space Telescope} (HST) imaging of a Local Group dwarf
galaxy, which we will denote as observational profile B. In particular,
we utilized the photometry from \citet{hid09}
using HST imaging of the dwarf transition-type galaxy, Phoenix,
for which the data are deep enough to sample the RGB 6 magnitudes
fainter than the TRGB. From the photometry provided by S. Hidalgo (private
communication; 2013), we fit functions of the form shown in Equation
(\ref{eq:err}) and (\ref{eq:compl}) to the photometric errors and
completeness. We note that \citet{bar04} utilized the same error
profile for both V and I, and we do the same here, as we found there to
be little difference between them. We therefore use
the fit from the I-band photometric errors. The coefficients
for profile B are also listed in Table \ref{tbl:err} for comparison. Typical errors on
the RGB for Profile B are $\sim$0.01 mag.  Both error and completeness profiles
are plotted for comparison in Figure \ref{fig:err}. Finally, in order
to realistically simulate the density of stars on the CMD for a system like 
a dwarf galaxy, we scale each synthetic CMD such that the number of stars 
at M$_{\mathrm I} = -3.0 \pm 0.1$ mag is $\sim$90. A typical CMD obtained using this
  profile contains $\sim$3500 stars brighter than M$_{\mathrm V}$ = 2 mag.
\begin{deluxetable}{ccccc}
\tablecaption{Coefficients for different observational profiles.\label{tbl:err}}
\tablewidth{0pt}
\tablehead{
\colhead{Profile} & \colhead{$\kappa$} &
\colhead{$\tau$} & \colhead{$\alpha$} &
\colhead{M$_0$}
}
\startdata
A & 0.11 & 0.39 & 2.0 & 1.0\\
B & 0.02 & 0.67 & 1.18 & 4.57\\
\enddata
\end{deluxetable}

\subsection{Metallicity Calculation}
\label{sub:met}
In an effort to derive metallicities from these synthetic CMDs in a
manner similar to what has been done in past work, we chose to
calculate the metallicity of each individual star on the RGB by
interpolating among a grid of theoretical isochrones. Our procedure is
modeled after that of \citet{sar05}, and we utilize the same
Interactive Data Language (IDL) code to perform the interpolations. To
summarize the procedure, the isochrones represent a three-dimensional
grid of M$_{\mathrm I}$, (V-I)$_0$, and metal abundance for stars of a given age. The IDL
routine \emph{trigrid} then performs the interpolation for each star
on the grid within a specified magnitude range. To estimate the errors
in metallicity arising from the presence of photometric errors,
we
repeated the interpolation to account for photometric error in
magnitude and color for each star, and
the sum of these deviations in quadrature is assigned as the error in
metallicity. We choose not to extrapolate beyond the grid since we found this to
result in highly uncertain metallicities. Therefore, if any
star falls off of the isochrone grid, either initially or after
scattering to account for the photometric errors, we discard it from
the final MDF.

Since the standard approach is to assume an old age for
most systems to which this technique is applied, we use isochrones for
an age of 12 Gyr as our fiducial grid. For systems older than $\sim$10 Gyr,
the interpolated metallicities from RGB stars do not depend
sensitively on the age of the isochrones used \citep{sar05}. Most
previous work with this method tends to exclude stars fainter than
some magnitude limit in order to remove contamination from stars on
the horizontal branch (HB) and red clump (RC). We therefore limit our interpolations to
magnitudes brighter than M$_{\mathrm{I}} = -1.2$ mag. Finally, to
obtain a singular estimate for the average metallicity, we construct a
generalized histogram of the MDF (i.e. the Gaussian smoothed MDF using
errors returned from the interpolation algorithm) and take the peak of
this distribution as the representative metallicity. We found this to
be more robust against skewing effects inherent in this method (see
Section \ref{sub:ssp}) than the
mean or median. For our analysis, the peak of the input MDFs are
calculated in a similar manner, with arbitrarily low errors assigned
to construct the distribution. Confidence intervals are calculated using a standard
bootstrapping technique, and the resulting reported errors correspond
to 99\% confidence intervals, or 3$\sigma$ uncertainties.

\section{Results}
\label{sec:res}
\subsection{A Simple Stellar Population: P1}
\label{sub:ssp}
\begin{figure}
\epsscale{1.2}
\plotone{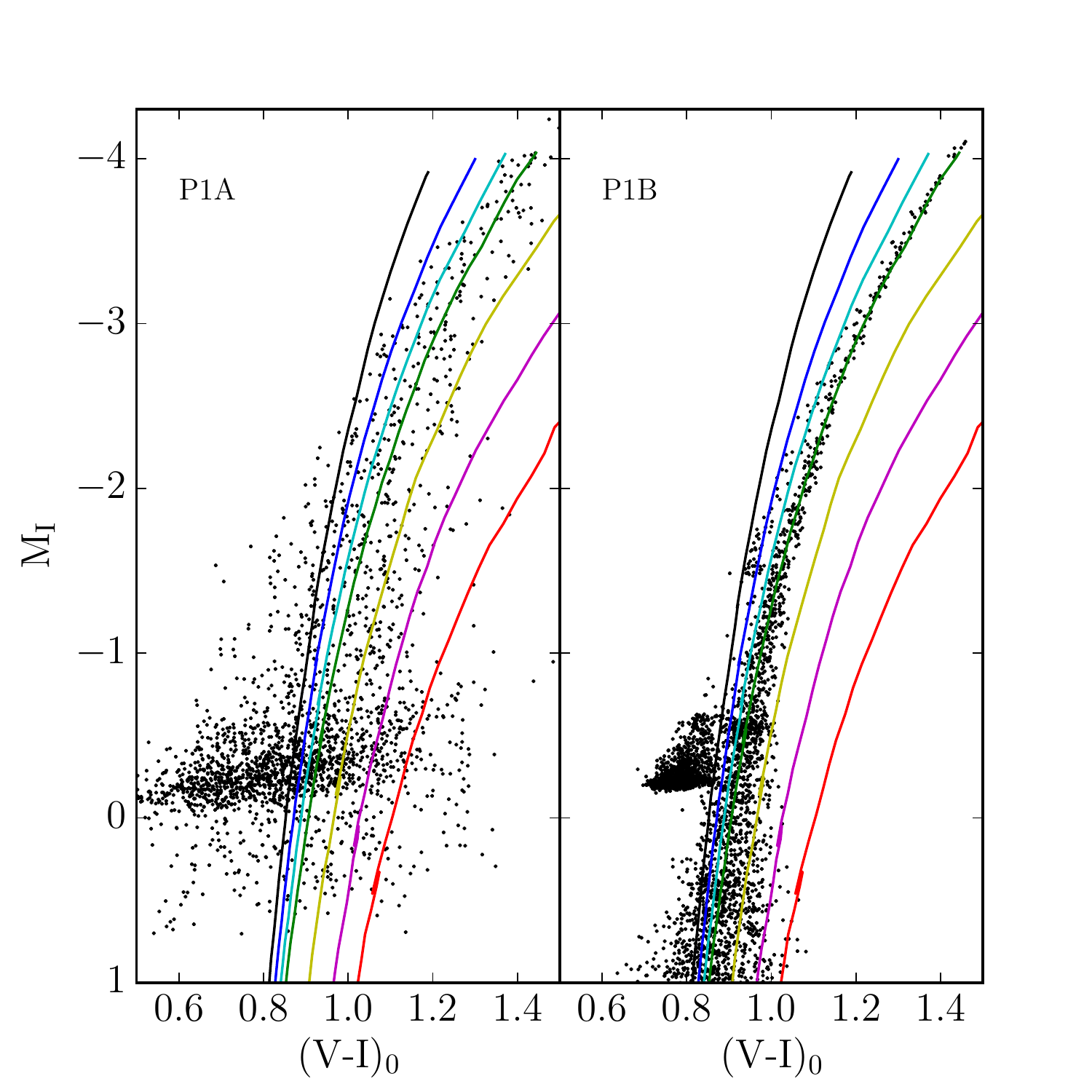}
\caption{\small{\emph{Left:} The M$_{\mathrm{I}}$, (V-I)$_0$ CMD for model
  1A. The model grid using the
isochrones of \citet{gir02} for an
  age of 12 Gyr and metallicities of, from left to right, $Z = 0.0001$, 0.0003, 0.0006, 0.001,
  0.002, 0.004, and 0.008 are overplotted for reference. Both the AGB
  and RC stars are strongly blended with the RGB due to the
  photometric errors. \emph{Right:} The CMD for model 1B. In CMD 1B, both the
  AGB and RC are apparent as distinct from the RGB, although some of
  these stars still contaminate the model grid of
  isochrones.\label{fig:1ab}}}
\end{figure}

We first synthesized the CMD of a simple population with one age of 12 Gyr,
one metallicity of $Z = 0.001$ ([M/H] = -1.28 dex), and no binaries, which
we will refer to as P1. We apply both observational
profiles to this synthetic CMD, to which we will refer to as models 1A and 1B (see Section
\ref{sub:err} for a description of the observational profiles). The CMDs
are plotted in Figure \ref{fig:1ab}. 

As was mentioned previously, studies that employ interpolation on
an isochrone grid typically limit the interpolation to stars brighter
than some magnitude limit to prevent contamination of RC and HB stars on the
model grid. Since a significant fraction of the AGB
occupies similar magnitudes as the bright RGB, these stars will generally provide some contamination if present within a
stellar population. Additionally,
fainter stars have larger photometric
errors, so at some point including fainter stars in the RGB may
only act to add noise to interpolated MDFs. To further investigate
how these factors affect the resultant MDF, we ran a
series of interpolations on these two CMDs each time changing the faint
magnitude limit, M$_{\mathrm I, f}$, above which the interpolation
was performed. In this way, we tested what magnitude range yields the
most accurate and precise metallicity estimate. We use
isochrones for an age of 12 Gyr and metallicities of $Z = 0.0001$, $Z
= 0.0003$, $Z = 0.0006$, $Z = 0.001$, $Z = 0.002$, $Z = 0.004$, and $Z
= 0.008$. The results from this set of interpolations are presented in Table
\ref{tbl:1ab} where the peak metallicity and its difference from the
input peak are presented for each case. 
\begin{deluxetable}{cccc}
\tablecaption{Results from varying the faint limit for the
  interpolation.\label{tbl:1ab}}
\tablewidth{0pt}
\tablehead{
\colhead{Model} & \colhead{M$_{\mathrm I,f}$ (mag)} &
\colhead{Peak [M/H]\tablenotemark{a} (dex)} 
& \colhead{$\Delta$[M/H] (dex)}}
\startdata
1A & -3.5 & $-1.25\pm 0.05$ & -0.03\\
1A & -3.0 & $-1.24\pm 0.06$ & -0.04\\
1A & -2.5 & $-1.23\pm 0.05$ & -0.05\\
1A & -2.0 & $-1.22\pm 0.06$ & -0.06\\
1A & -1.5 & $-1.21\pm 0.05$ & -0.07\\
1A & -1.2 & $-1.20\pm 0.05$ & -0.08\\
1B & -3.5 & $-1.29\pm 0.01$ & 0.01\\
1B & -3.0 & $-1.29\pm 0.01$ & 0.01\\
1B & -2.5 & $-1.29\pm 0.01$ & 0.01\\
1B & -2.0 & $-1.29\pm 0.01$ & 0.01\\
1B & -1.5 & $-1.29\pm 0.01$ & 0.01\\
1B & -1.2 & $-1.29\pm 0.01$ & 0.01\\
\enddata
\tablenotetext{a}{These results correspond to a simple stellar
  population with age 12 Gyr and [M/H] = -1.28 dex.}
\tablenotetext{b}{Quoted uncertainties represent bootstrapped 99\% confidence intervals.}
\end{deluxetable}
\begin{figure*}
\epsscale{2.0}
\plotone{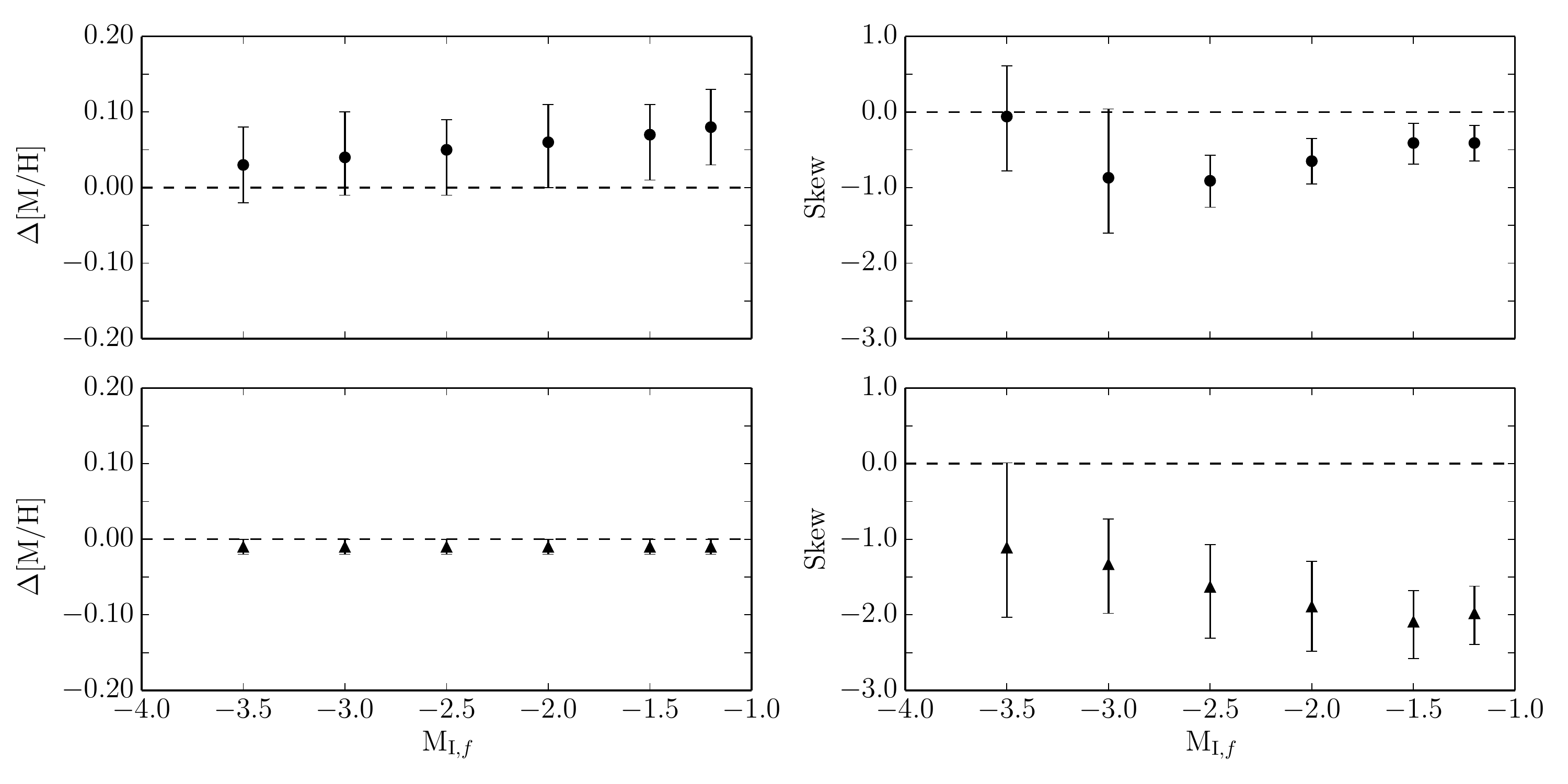}
\caption{\small{\emph{Left:} Difference between peak metallicity estimates from isochrone
  interpolation and the input metallicity for Models 1A (circles) and 1B
  (triangles) as a function of the adopted threshold magnitude,} M$_{\mathrm I, f}$. \emph{Right:} Skew of interpolated MDFs as a function
  of the adopted threshold magnitude, M$_{\mathrm I, f}$. The dashed lines represent the input MDF values. \emph{Top:} Interpolations for Model 1A. \emph{Bottom:}
  Interpolations for Model 1B. The error bars represent 99\%
  confidence intervals from a standard percentile bootstrap analysis.\label{fig:tbl1}}
\end{figure*}

Overall, it is clear that the interpolations produce
fairly accurate values for the peak metallicity, given that all are within 0.1
dex of the input values. We illustrate
this further in Figure \ref{fig:tbl1} where we have plotted the
difference between the input and output peak metallicities against
M$_{\mathrm{I}, f}$. 

For model 1B, the interpolation proves very reliable owing to the
smaller photometric errors (\ref{fig:tbl1}; bottom left), yielding peak metallicities
which are both accurate and precise to within 0.01
dex. It seems that no matter where the interpolation is limited, the
peak metallicities in the case of the small error profile yield
systematically lower metallicities. The skew of this MDF, also
illustrated in Figure \ref{fig:tbl1}, is consistent with skewness
towards metal-poor values. 

While this systematic offset in metallicity is
within the errors, we note that one possible reason for systematically
metal-poor values could be the AGB stars contaminating the grid at all
magnitudes. Another possibility is that the asymmetric nature of the
isochrone grid in the magnitude-color plane (i.e. the nonlinear
dependence of the color of the RGB on [M/H]) produces this negative
skew. Since the grid is asymmetric but the photometric errors are
applied symmetrically, it follows that this effect could lead to some amount
of artificial skewing. In order to further investigate these effects, we repeated the
interpolations on P1 with the same observational profiles applied,
however this time we excluded AGB stars from the final CMDs. This was
achieved by applying cuts in initial mass, magnitude, and color such
that stars more evolved than stars at the TRGB were filtered out.
  Note that the cuts in magnitude
and color were only required due to the small but nonzero range in the
ages of the stars in this simple stellar population resulting from the
interpolation in time of the SFR by IAC-STAR. We
then performed the interpolation down to magnitudes of M$_{\mathrm{I,
    f}}$ = -1.2 mag. The resulting MDFs did not show any significant
differences in peak metallicity or skew, forcing us to conclude that
the AGB does not provide a significant source of contamination for an old,
simple stellar population. Consequently, the asymmetry of the
isochrone grid remains the only valid explanation, and we conclude
that this asymmetry
does provide a significant source of artificial negative skew to an
interpolated MDF.

On the other hand, the model 1A yields systematically more
\emph{metal-rich} peaks than the input (Figure \ref{fig:tbl1}; top left). Additionally, 
this discrepancy grows as more faint RGB stars are included in the
interpolation. While these differences are mostly within the errors, its 
systematic nature could be of concern. We
attribute this behavior to the interplay between the photometric errors and the
isochrone grid. More specifically, the error profile for
model 1A is such that a significant number of stars are observed off
of the isochrone grid. This effect becomes more prominent for fainter
stars since they have larger photometric errors, and the effect is
asymmetric with respect to color. That is to say, the grid extends
further in the redward direction than in the blueward direction. Thus,
even though stars are scattered symmetrically in color due to the
photometric errors, a greater number of stars will 'fall off' the
model grid at the blue end. Simultaneously, stars that are scattered
to redder colors are more likely to stay on the grid. This appears to
diminish the skew of the MDF and move the peak to more metal-rich
values. Once again,
the skewness for model 1A (Figure \ref{fig:tbl1}; top right) seem to
corroborate this behavior. The interpolated MDFs of model 1A appear systematically less
skewed than those of model 1B. 

Considering these effects, we now turn to determining the optimal cutoff
for a metallicity interpolation on the RGB. Since we have determined
AGB contamination to be negligible in an old stellar population, the fundamental
competition governing the magnitude cutoff is between Poisson
statistics and photometric errors. On the one hand, excluding more stars at
fainter magnitudes reduces the number of stars in the sample,
potentially leading to a metallicity estimate based on a prohibitively small sample
of stars. On the other hand, including stars at fainter magnitudes
allows stars with increasingly uncertain photometry, and thus
metallicity, into the sample. This may lead, for example, to the peak
metallicity discrepancy observed for model 1A. Thus, it seems that when
the size of the photometric errors becomes comparable to the extent of
the isochrone grid, interpolated metallicities may be subject to
systematic errors. 

To compromise between these two competing effects, we examined how the
number of stars in the interpolation sample changed with
M$_{\mathrm{I},f}$. Moving the limit from M$_{\mathrm{I},f}$ = -1.2 mag to
M$_{\mathrm{I},f}$ = -1.5 mag results in losses of $\sim$14\% (1A) and
$\sim$20\% (1B) of the stars above M$_{\mathrm I}$ = -1.2 mag ( note
the difference in fractional losses is due to the completeness profiles). Additionally, the interpolated MDFs are nearly
identical in each case, with the only exception being a more accurate
value for the peak metallicity in model 1A (Figure \ref{fig:tbl1}; top
left). Making the cutoff brighter at M$_{\mathrm{I},f}$ = -2.0
mag resulted in losses of 40-50\% of the stars above M$_{\mathrm I}$ = -1.2 mag
with no significant increase in the accuracy of the interpolated
metallicities. Under these considerations, we choose
M$_{\mathrm{I},f}$ = -1.5 mag as our optimal cutoff and limit all future
interpolations to stars brighter than this magnitude. 

An anonymous referee suggested an alternative method to derive
the peak metallicity of a population by constructing a
fiducial RGB sequence from the photometry of the observed RGB. To
accomplish this, we binned all of the stars brighter than the optimal
cutoff, M$_{\mathrm{I},f}$ = -1.5 mag, into bins of 0.1 mag and then
calculated the median (V-I)$_0$ color within each bin. The
subsequent color paired with the midpoint of each magnitude bin served
as our fiducial RGB which was input into the metallicity interpolation
routine. We applied this method to both models 1A and 1B,
and and the resulting
differences between the input and output peaks in each case was 0.03
dex for model 1A and 0.01 dex for 1B. Comparing these $\Delta$[M/H]
with those listed in Table \ref{tbl:1ab}, it appears that this
alternative method works well and recovers similarly accurate peaks. However, since this method removes
information about the shape of the MDF, we shall continue to use the
original, star by star interpolation to construct MDFs for the
remainder of the paper.

Finally, we investigated how the metallicity of the simple stellar
population affected the resulting metallicity interpolation. To this
end, we synthesized a population whose properties were identical to
P1, but having a metallicity of $Z = 0.0076$ ([M/H] = -0.4 dex). We
performed the interpolation on the same grid in a similar manner to
P1, and the errors in the peak metallicity amounted to $\sim$0.03 dex
in [M/H]. Therefore, we conclude that the results of this
interpolation method to be relatively insensitive to the bulk
metallicity of the stellar population.

We conclude this section with some summarizing remarks. For an old, simple
stellar population, isochrone interpolation on the RGB works very well to
retrieve the peak metal abundance, with metallicity errors less
than 0.1 dex for the worst case. In cases where the photometric
errors are large with respect to the size of the isochrone grid,
exemplified in model 1A, the interpolated MDF and therefore peak
metallicity are sensitive to the shape of the isochrone grid. For
cases where the errors are small with respect to the model grid (1B),
the MDF and its peak are accurately recovered to within 0.01
dex. We determine the best compromise between Poisson errors
and photometric errors of an interpolated MDF to be to limit the
interpolation to stars brighter than M$_I$=-1.5 mag. Finally, we
conclude that this method is insensitive to the bulk metallicity of
the population.

\subsection{The Effects of a Metallicity Spread: P2}
\label{sub:metspread}
The limit of a simple stellar population with one age and one
metallicity (i.e. [Fe/H]) is approximately valid for systems such as globular
clusters (the effects of multiple populations
and enhancement differences in individual globular clusters is beyond the
scope of this work). However more complex systems, such as dwarf
galaxies, require a more complex treatment. In particular, these
systems typically show extended SFHs resulting in significant chemical
evolution over time. Additionally, inherent metal abundance variations
within a stellar population will also affect a metallicity
determination. As a simple exploration of the effects of a metallicity
spread, we created a synthetic CMD whose
properties are identical to P1 (Section \ref{sub:ssp}), with the
only difference being a spread in the stellar metallicities. To
accomplish this, we utilize a feature of IAC-STAR which allows the
user to provide an upper and lower bound for the input AMR. IAC-STAR
then calculates the metallicity of a star formed at any instant in
time by randomly sampling a Gaussian bounded by these two AMRs at that
time. 

\begin{figure*}
\epsscale{2.0}
\plotone{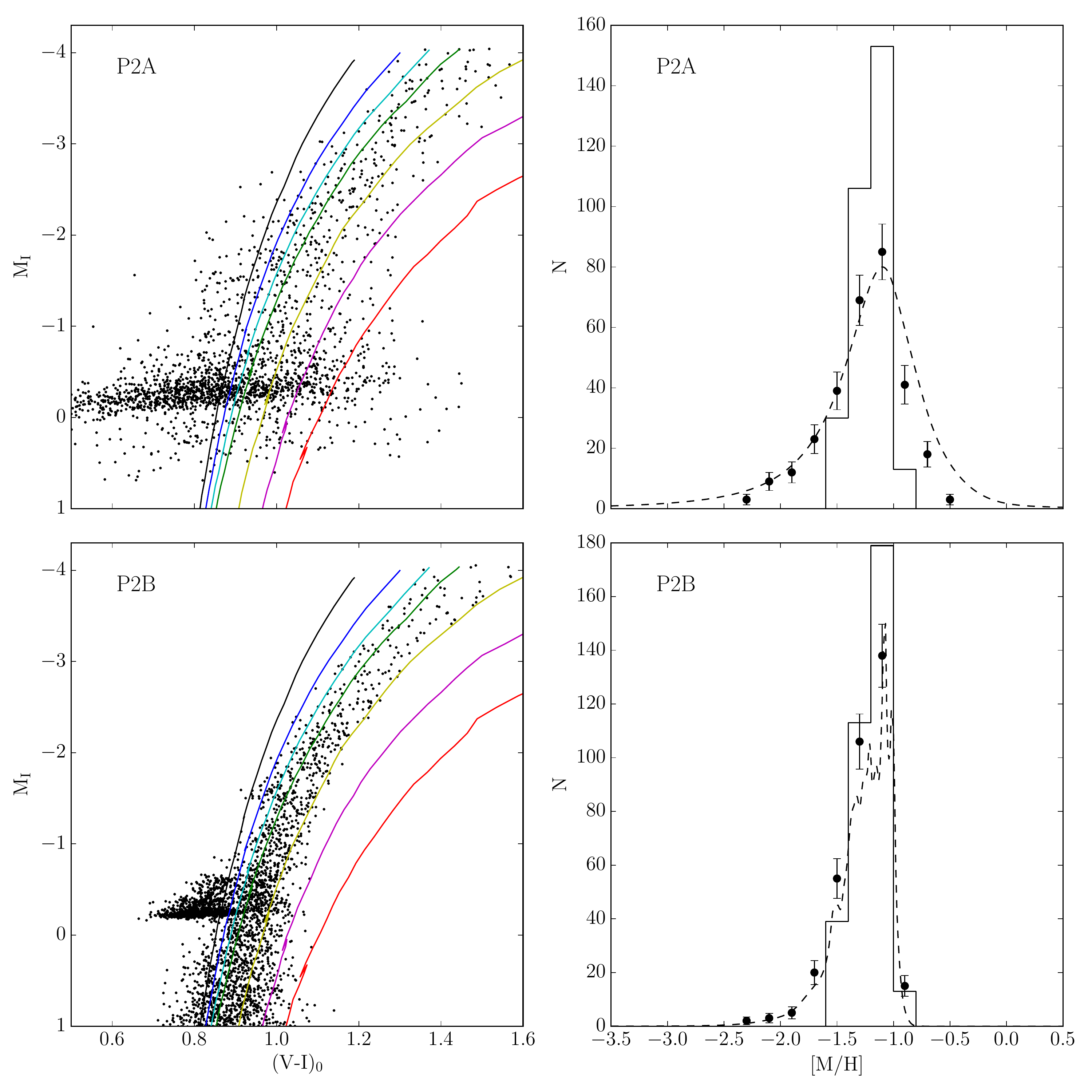}
\caption{\small{\emph{Left:} The M$_{\mathrm{I}}$, (V-I)$_0$ CMD for models
  2A (\emph{top}) and 2B (\emph{bottom}). The model grid using the
isochrones of \citet{gir02} for an
  age of 12 Gyr and metallicities of $Z = 0.0001$, 0.0003, 0.0006, 0.001,
  0.002, 0.004, and 0.008 are overplotted for reference.\emph{Right:} The
  interpolated binned (points) and Gaussian smoothed (dashed line) MDFs for
  each corresponding model CMD. The error bars in the binned MDFs are
  1$\sigma$ Poisson uncertainties. The solid bar plots represent the
  input MDFs.\label{fig:2ab}}}
\end{figure*}

In this
case, we set the lower AMR to a metallicity of $Z = 0.0006$
([M/H] = -1.51 dex) and the upper AMR to a metallicity of $Z = 0.002$ ([M/H]
= -0.98 dex), each constant in time, and we will henceforth refer to this as P2. The
resulting input MDF is characterized by a mean metallicity of [M/H]
= -1.20$\pm$0.15 dex, where the spread is calculated as one standard
deviation. As with P1,
we apply both observational profiles to this population and refer to
these two CMDs as models 2A and 2B. We then performed the metallicity
interpolation using the same grid from Section \ref{sub:ssp}. The
resulting model CMDs and MDFs are plotted together in Figure
\ref{fig:2ab}. 

The peak interpolated metallicities for these model CMDs are [M/H] =
-1.10$^{+0.06}_{-0.06}$ dex for 2A and [M/H] = -1.07$^{+0.15}_{-0.07}$ dex
for 2B, and they differ from their input peaks by 0 dex and
-0.04 dex respectively. Thus, each peak is again recovered well within
the errors. Interestingly, model 2A has its peak recovered more
accurately than model 2B. Although the discrepancy is within
the errors, it is interesting to note that the skewness of MDF 2B is
significantly more negative than for its input MDF. This is not true
for MDF 2A. Similar behavior is observed for the simple stellar
population in Section \ref{sub:ssp} (Figure \ref{fig:tbl1}; right
column). Since the only difference between the two model CMDs is the
observational profile applied, it seems that the photometric errors
must be responsible for this. We interpret this to be related to the
phenomenon previously observed and discussed in Section
\ref{sub:ssp}, namely the interaction of the photometric errors with
the isochrone grid. Thus, even though MDF 2A suffers from more noise
resultant from the larger photometric errors, it is less negatively
skewed than MDF 2B because more stars are scattered off of the blue,
metal-poor end of the grid than the red, metal-rich end. We emphasize
that this systematic effect results in recovered peak metallicity for
MDF 2A lying near the center of the isochrone grid and is not a
reflection of higher accuracy. This
illustrates the importance of ensuring that the photometric errors in
a CMD upon which metallicity interpolation will be applied are small
in comparison to the extent of the isochrone grid. In this case
however, this effect is small. 

To add to this, the error in the
peak for 2A is significantly smaller than for 2B, even though the
former suffers from larger
photometric errors than the latter. To investigate the cause of this,
we examined the bootstrapped distributions of the MDF peak for both
model CMDs. This revealed a multimodal distribution of peak
metallicities for MDF 2B. Specifically, in addition to the true peak
at [M/H] = -1.07 dex, there also exist less prominent but still
significant peaks at [M/H]$\sim$-1.22 dex and [M/H]$\sim$-1.03
dex. Careful examination of the generalized histogram of MDF 2B
(Figure \ref{fig:2ab}; bottom right)
reveals these secondary peaks. These secondary peaks act to add
uncertainty to the primary peak via our bootstrapping technique. It is
therefore likely that we are over-estimating the errors in this case
by using the 3$\sigma$ errors. On the other
hand, the larger photometric errors in CMD 2A smear these peaks out
beyond recognition such that they do not present themselves in the
bootstrapped distribution of the MDF peak. Taking into consideration
all of the effects discussed in this subsection, we conclude that the peak interpolated
metallicity is relatively insensitive to the effects of an
inherent metallicity spread of order $\sigma \sim 0.15$ dex. 

We conclude this subsection by noting that overall, the shape of
  the interpolated MDF 2B (\ref{fig:2ab}; bottom left, points) closely
  matches that of the input MDF (\ref{fig:2ab}; bottom left, solid
  line). The same is not true for MDF 2A (\ref{fig:2ab}). We note that
the input MDFs for the two model CMDs differ slightly due to Poisson
statistical effects introduced during the application of the
completeness profile.

\subsection{The Effects of an Age Spread: P3}
\label{sub:agespread}
\begin{figure*}
\epsscale{2.0}
\plotone{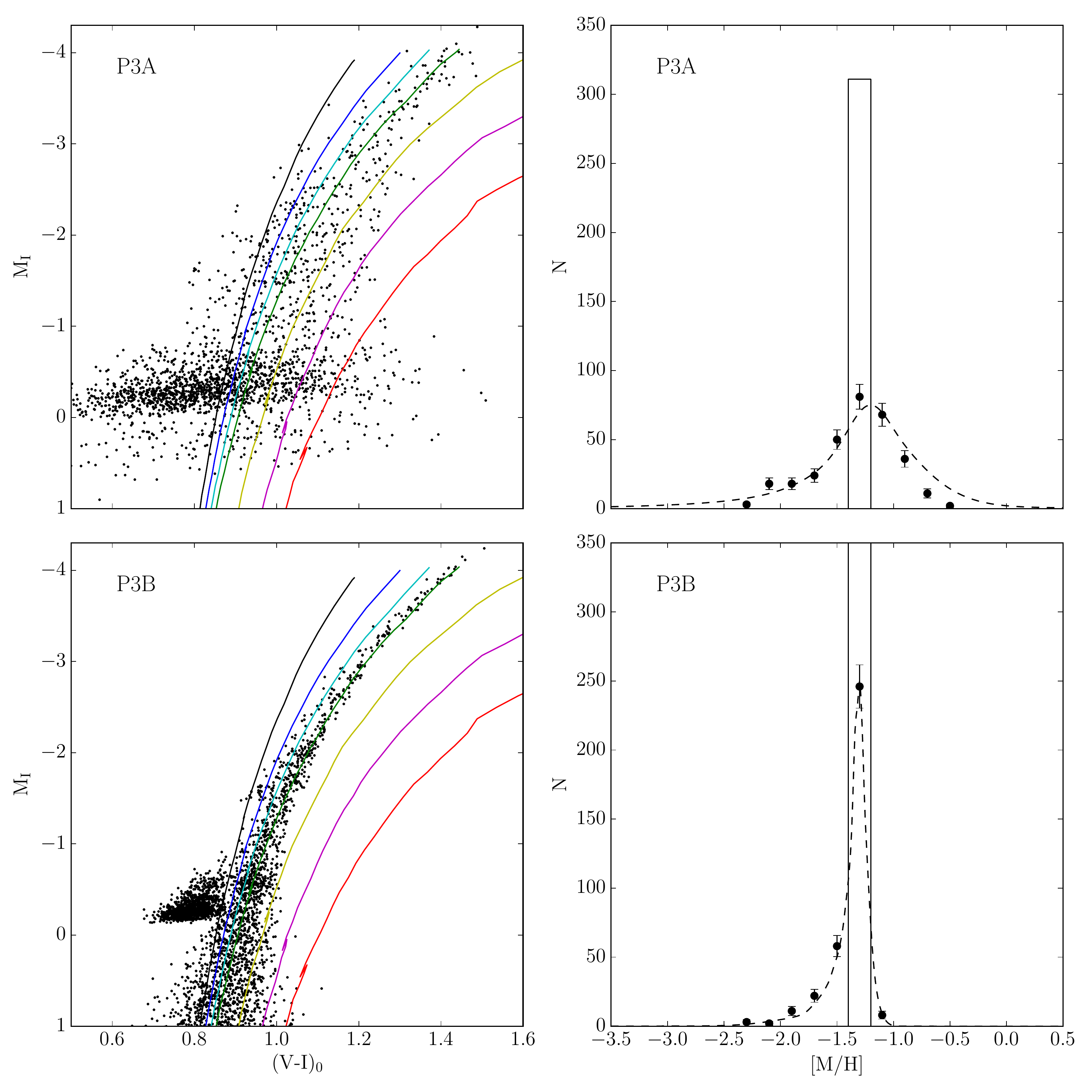}
\caption{\small{\emph{Left:} The M$_{\mathrm{I}}$, (V-I)$_0$ CMD for models
  3A (\emph{top}) and 3B (\emph{bottom}). The model grid using the
isochrones of \citet{gir02} for an
  age of 12 Gyr and metallicities of $Z = 0.0001$, 0.0003, 0.0006, 0.001,
  0.002, 0.004, and 0.008 are overplotted for reference.\emph{Right:} The
  interpolated binned (points) and Gaussian smoothed (dashed line) MDFs for
  each corresponding model CMD. The error bars in the binned MDFs are
  1$\sigma$ Poisson uncertainties. The solid bar plots represent the
  input MDFs.\label{fig:3ab}}}
\end{figure*}

We now turn to exploring the effects of a spread in the stellar
ages. For this we synthesized a synthetic CMD whose properties are
similar to the simple stellar population, P1, but whose SFH consists
of a Gaussian with a mean age of 12 Gyr and a 1$\sigma$ width of 2
Gyr. We designate this population as P3, and again apply the two
observational profiles creating two models CMDs 3A and 3B. We
performed the interpolation as before, and the results are illustrated
in Figure \ref{fig:3ab}.

The peak interpolated metallicities in this case are [M/H] =
-1.20$^{+0.04}_{-0.05}$ dex for 3A and [M/H] = -1.30$^{+0.04}_{-0.02}$
dex for 3B. We note that these results are similar to that for the
simple stellar population P1 (see Section \ref{sub:ssp}). That is,
the peak metallicity is well-recovered in the case of smaller
photometric errors, albeit slightly metal-poor. The only effect of the age spread in this case is to
increase the scatter from $\sim$0.01 dex to a few hundredths of a dex. On
the other hand, the peak for 3A with its larger
photometric errors is again more \emph{metal-rich} than the input
metallicity. This seems to be the result of the interplay of the
photometric errors with the isochrone grid. In this case, the
uncertainty in the peak remains the same as for a simple stellar
population, indicating that the photometric errors wash out any signal
of the age spread in this case. We therefore conclude that
for an old age, the effects of an age spread of order $\sigma \sim 2$
Gyr do little to affect a
metallicity interpolation. The peak is recovered with
similar accuracy to a single-age population. The only observed effect is an increase in
the error of the peak for the small error case, resulting in
uncertainties still well below 0.1 dex in both cases.

\subsection{The Effects of an Age and Metallicity spread: P4}
\label{sub:metagespread}
\begin{figure*}
\epsscale{2.0}
\plotone{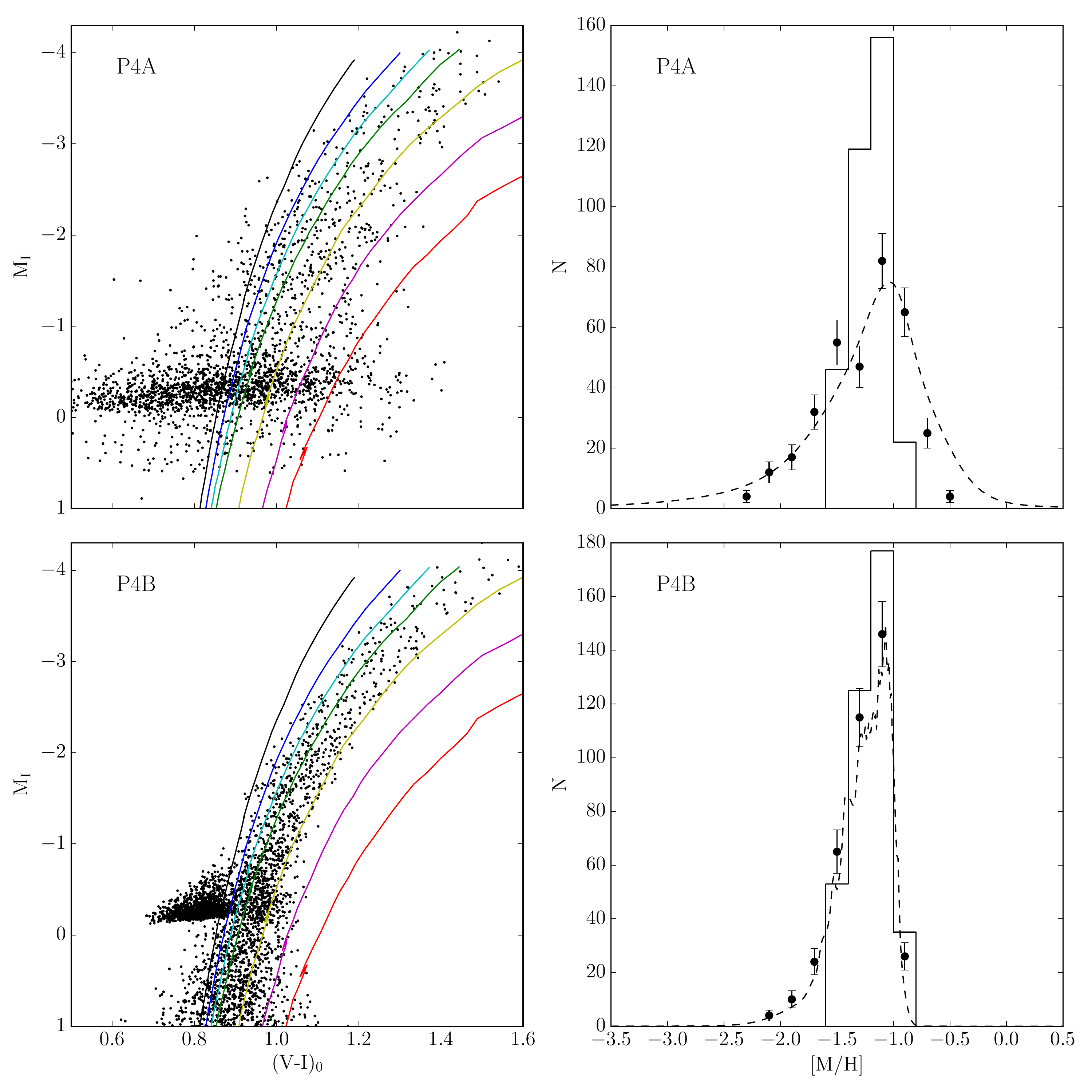}
\caption{\small{\emph{Left:} The M$_{\mathrm{I}}$, (V-I)$_0$ CMD for models
  4A (\emph{top}) and 4B (\emph{bottom}). The model grid using the
isochrones of \citet{gir02} for an
  age of 12 Gyr and metallicities of $Z = 0.0001$, 0.0003, 0.0006, 0.001,
  0.002, 0.004, and 0.008 are overplotted for reference.\emph{Right:} The
  interpolated binned (points) and Gaussian smoothed (dashed line) MDFs for
  each corresponding model CMD. The error bars in the binned MDFs are
  1$\sigma$ Poisson uncertainties. The solid bar plots represent the
  input MDFs.\label{fig:4ab}}}
\end{figure*}

We now add complexity to the synthetic CMD by inputting spreads in
both metallicity and age. For this case, we essentially combine the
AMR from P2 and the SFH from P3 and create an old population with mean
age 12 Gyr, age spread of 2 Gyr, mean metallicity of [M/H]$= -1.20$
dex, and metallicity spread of 0.15
dex. This population is designated P4, and applying the observational
profiles as before we have two model CMDs 4A and 4B. Performing the
interpolations in the same way as with the previous models, we construct the MDFs along with their
corresponding CMDs in Figure \ref{fig:4ab}.

The peak interpolated metallicities for models 4A and 4B are [M/H] =
-1.04$^{+0.08}_{-0.06}$ dex and [M/H] = -1.07$^{+0.19}_{-0.05}$ dex
respectively. The peaks for both input MDFs are near [M/H] = -1.00 dex, so
both peaks are recovered within their respective confidence
intervals. One may note that the quoted uncertainties for these values
are similar to those of models 2A and 2B, indicating that the
metallicity spread dominates over the age spread in this case. Again,
here it appears that we may be overestimating the errors in the case
with the smaller photometric errors. The same mechanism discussed for
model 2B in
Section \ref{sub:metspread} seems to be at work for model 4B, namely
the asymmetry of the isochrones in the CMDs and how this affects the metallicity
interpolation. Aside
from this, the peak metallicities are well-recovered in this case with
both an age and metallicity spread. 

\subsection{The Effects of an Age Error}
\label{sub:ageerr}
\begin{figure*}
\includegraphics[scale=0.9, angle=90]{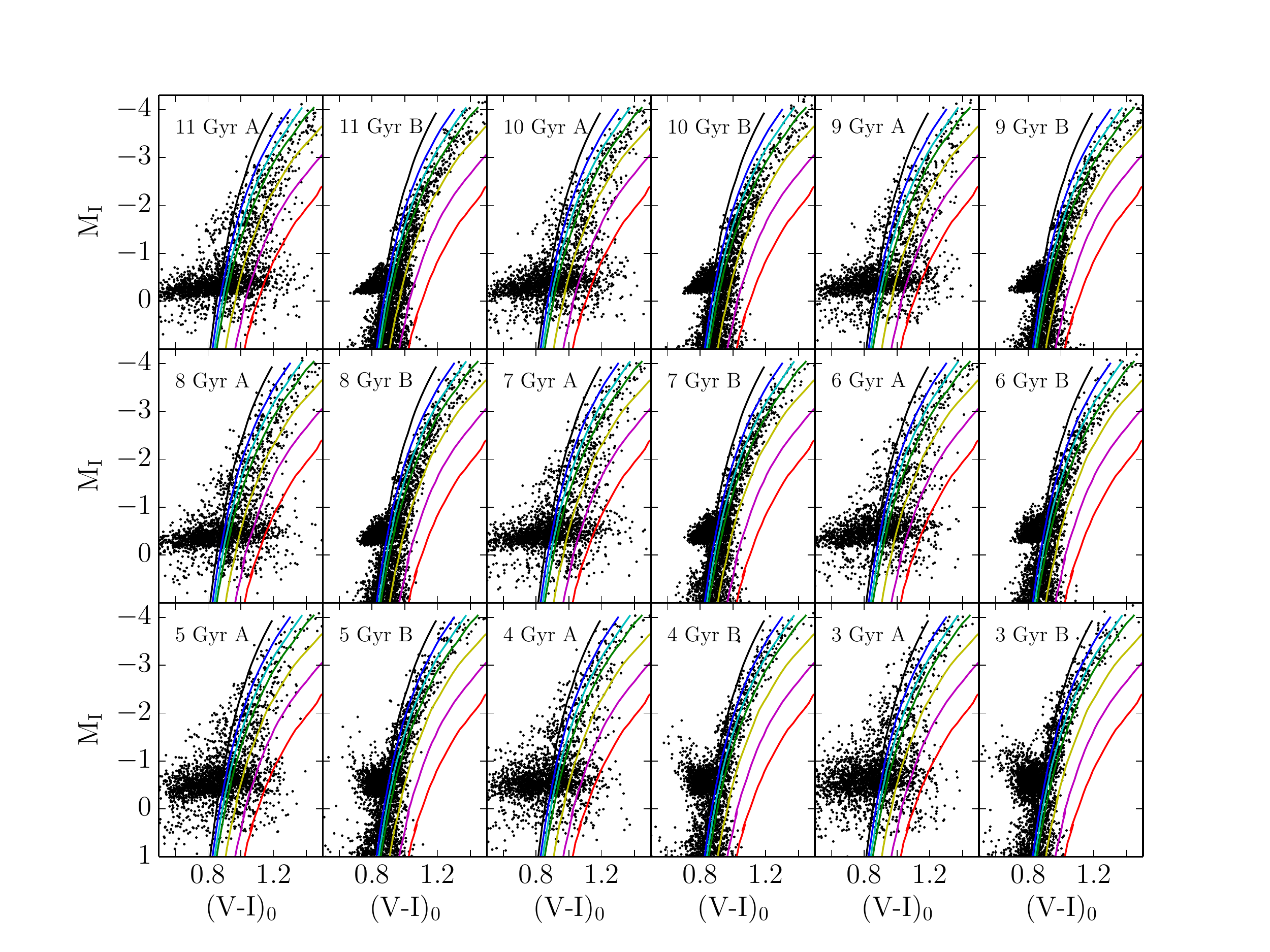}
\caption{\small{The model CMDs used in investigating the effects of
    an age error on a metallicity interpolation.\label{fig:agecmds}}}
\end{figure*}

In the absence of an independent measure of the age of a stellar
system, it is also interesting to investigate how a systematic error
in the assumed age will affect a metallicity determination using the
RGB. As was discussed above, for an old population
($\gtrsim$10 Gyr) the RGB is relatively insensitive to the exact
age. However, for younger populations this does not hold. To explore
this further, we generated a series of synthetic CMDs each with the
same properties as P4, except changing the mean age of the stars by 1
Gyr from 11 Gyr down to 3 Gyr. We then applied the observational
profiles and interpolated in metallicity on the 12 Gyr isochrone grid
to observe how the peak metallicity varies with a fixed adopted age of
12 Gyr. 
Each of these model CMDs is displayed in Figure \ref{fig:agecmds}.
\begin{figure}
\epsscale{1.2}
\plotone{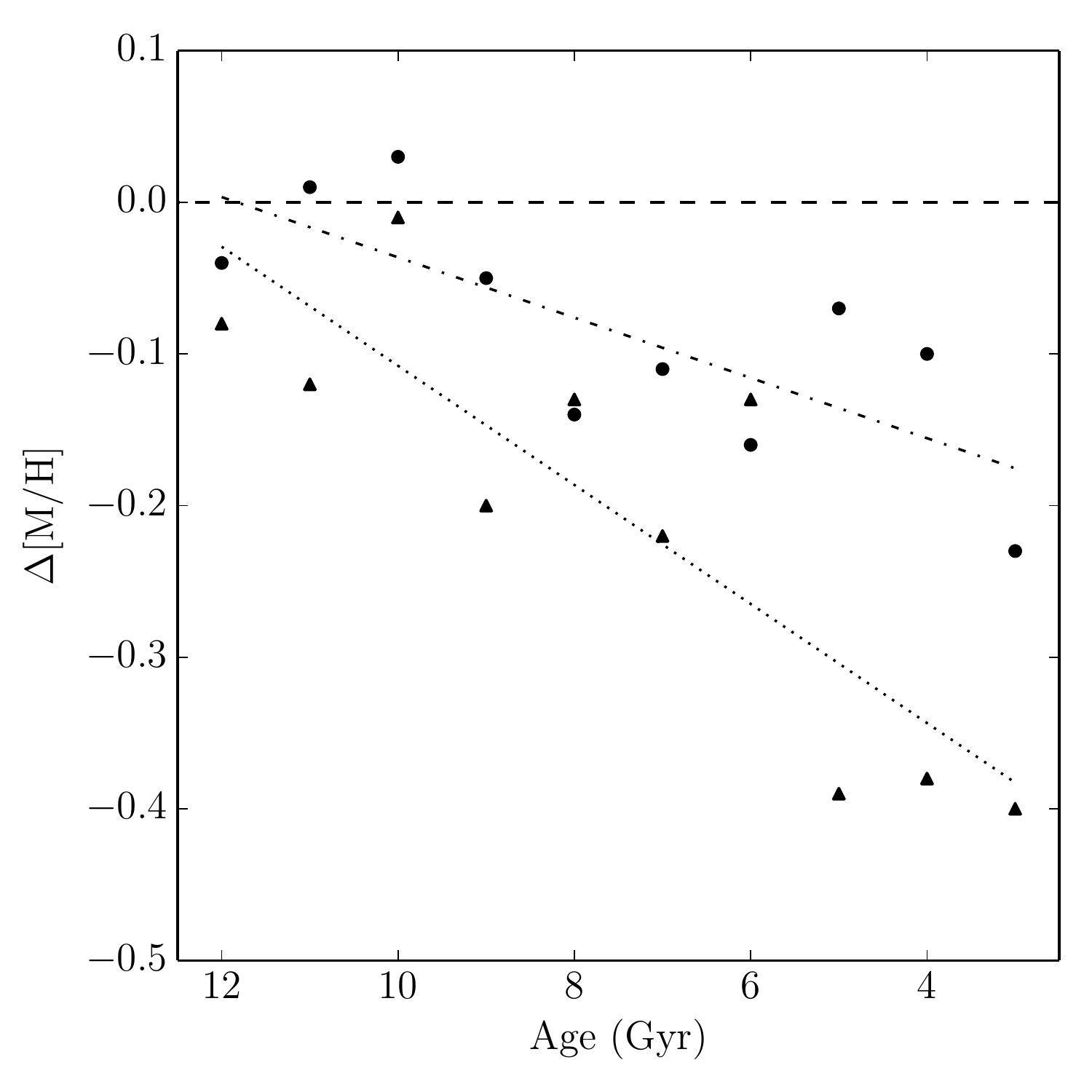}
\caption{\small{Peak metallicity discrepancy as a function of the mean
    age of a synthetic CMD. Circles represent CMDs modeled with
    observational profile A, while triangles are CMDs modeled with
    profile B. The dashed-dot and dotted lines show the linear fits
    for the two profiles A and B, respectively.\label{fig:agemet}}}
\end{figure}

Figure \ref{fig:agemet} shows how the metallicity discrepancy changes
as a function of mean population age for model CMDs using the
observational profiles A (circles) and B (triangles). Here, $\Delta$[M/H] is in the
sense [M/H]$_{\mathrm{out}}$ -[M/H]$_{\mathrm{in}}$. Figure
\ref{fig:agemet} shows a clear trend in which populations younger than the
isochrones in the model grid produce systematically more metal-poor
metallicities. In accordance with previous studies, this effect is
small ($\lesssim$0.1 dex) in populations older than 10 Gyr but may
produce an error of nearly half a dex for considerably younger
populations. Linear fits to the data resulted in the parameters presented
in Table \ref{tbl:agelin}. The relations are incompatible
with each other considering the quoted uncertainties. It seems that
CMDs with smaller photometric errors along the RGB are more sensitive
to age differences in their interpolated MDFs. Part of the decreased
sensitivity for profile A may be the coupling between the isochrone
grid and larger photometric errors discussed in Section
\ref{sub:ssp}. 
\begin{deluxetable}{cccc}
\tablecaption{Linear fit parameters for metallicity error versus age
  offset experiments.\label{tbl:agelin}}
\tablewidth{0pt}
\tablehead{
\colhead{Profile} & \colhead{Slope\tablenotemark{a} (dex/Gyr)} &
\colhead{Zero-point\tablenotemark{a} (dex)} & \colhead{RMS scatter (dex)}
}
\startdata
A & 0.020$\pm$0.007 & -0.235$\pm$0.055 & 0.049\\
B & 0.039$\pm$0.010 & -0.501$\pm$0.079 & 0.069\\
\enddata
\tablenotetext{a}{Quoted uncertainties represent the square roots of
  the diagonal elements from the covariance matrix returned from the
  fitting routine.}
\end{deluxetable}

Of particular note for the small error profile
(\emph{triangles}) is the discontinuous jump in metallicity error from
6 to 5 Gyr. It seems that this age range defines where the stars
present on the interpolation grid become increasingly dependent on the
age of the system.

On the other hand, for the case with larger errors (\emph{circles}) we
observe systematically more accurate peak metallicities. While this is
certainly counter-intuitive, this is almost certainly a result of the
coupling between the photometric errors and the adopted isochrone
grid. As discussed in Section \ref{sub:ssp}, the large photometric
errors in profile A couple with the asymmetry of the isochrone grid to
scatter more stars off of the metal-poor end compared with the
metal-rich end of the grid. Thus, the seemingly more accurate values
for CMDs obtained applying profile A is more likely an artifact of the
interpolation procedure rather than a reflection of the true
accuracy. 

\subsection{Multiple Gaussian bursts: P5}
\label{sub:trigauss}
\begin{figure}
\epsscale{1.2}
\plotone{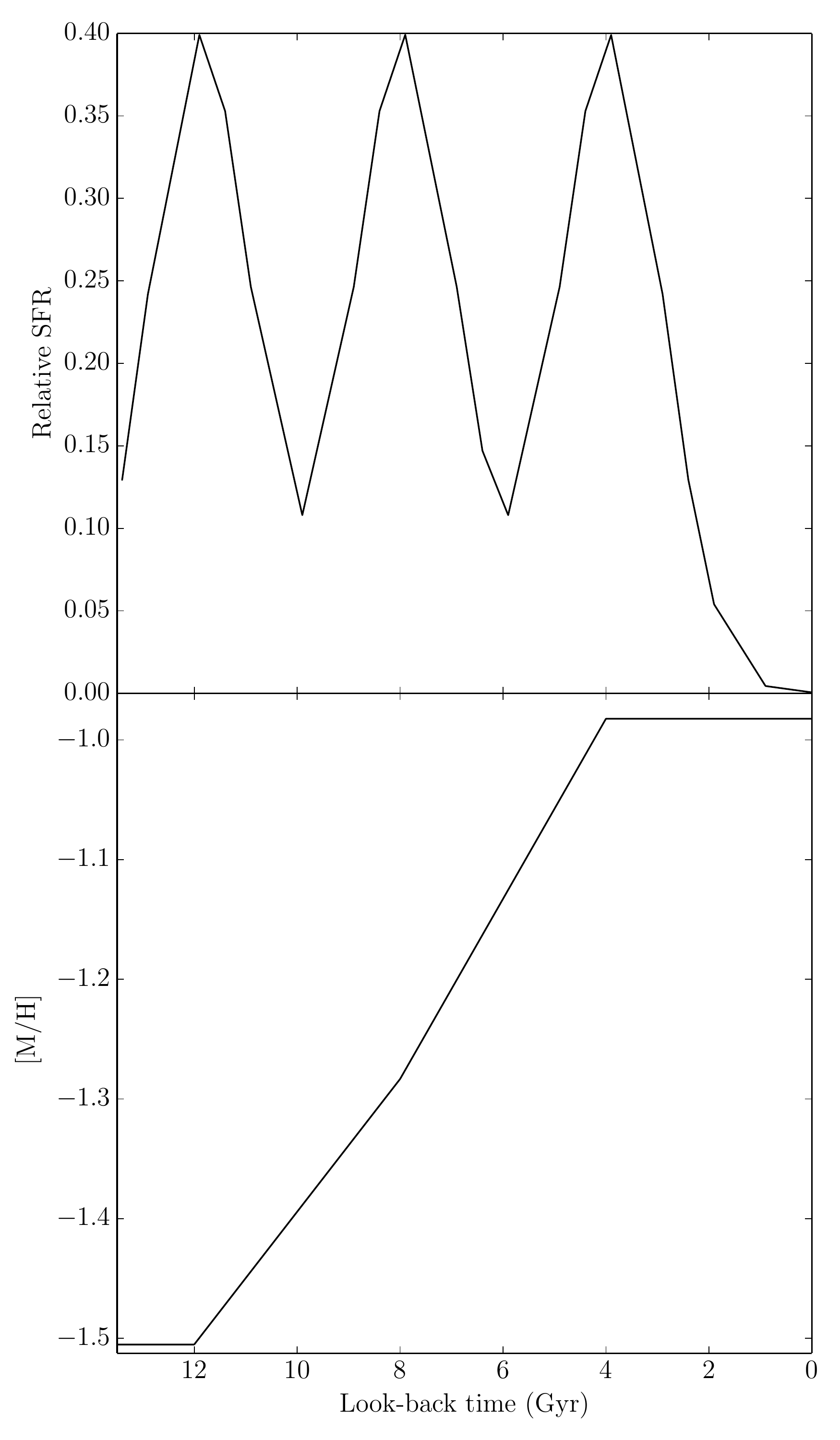}
\caption{\small{\emph{Top:} The input SFH for P5 consisting of 3
    Gaussian bursts of star formation. Note that IAC-STAR cannot take
    more than 20 nodes in time for the input SFH, hence the coarse
    resolution here. IAC-STAR does however interpolate amongst these nodes in time to
    achieve greater time resolution. \emph{Bottom:} The input AMR
    for P5.\label{fig:trigauss}}}
\end{figure}

\begin{figure*}
\epsscale{2.0}
\plotone{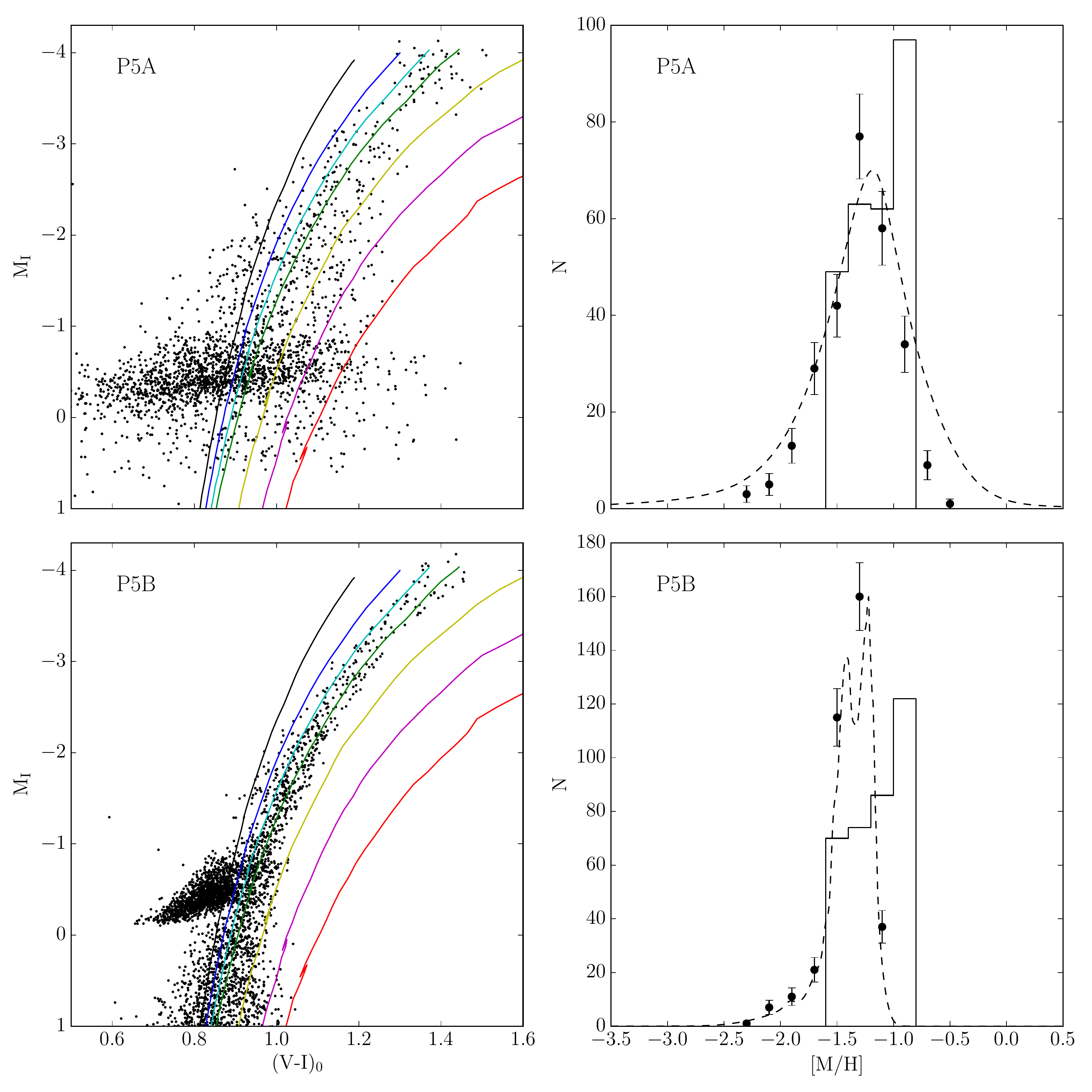}
\caption{\small{\emph{Left:} The M$_{\mathrm{I}}$, (V-I)$_0$ CMD for models
  5A (\emph{top}) and 5B (\emph{bottom}). The model grid using the
isochrones of \citet{gir02} for an
  age of 12 Gyr and metallicities of $Z = 0.0001$, 0.0003, 0.0006, 0.001,
  0.002, 0.004, and 0.008 are overplotted for reference.\emph{Right:} The
  interpolated binned (points) and Gaussian smoothed (dashed line) MDFs for
  each corresponding model CMD. The error bars in the binned MDFs are
  1$\sigma$ Poisson uncertainties. The solid bar plots represent the
  input MDFs.\label{fig:5ab}}}
\end{figure*}

We tested the effects of multiple epochs of star formation in a system
by simulating a stellar population with 3 distinct star formation
episodes. We achieved this by inputting a SFH into IAC-STAR which
consisted of three Gaussian bursts of star formation at ages of 12,
8, and 4 Gyr. All have 1$\sigma$ widths of 1 Gyr, and the AMR is
monotonically increasing between the peaks of star formation. Both the
SFH and AMR for this population, which we call P5, are illustrated in
Figure \ref{fig:trigauss}. Applying the observational profiles and
interpolating on the same isochrones as in previous sections yields
CMDs 5A and 5B. These CMDs and corresponding interpolated MDFs are
illustrated in Figure \ref{fig:5ab}.

The peak of the input metallicity lies at [M/H] = -0.98 dex. Since the
most recent star formation episode at 4 Gyr dominates the number of
stars currently present on the RGB, the MDF is peaked at this metal-rich
end. However, the interpolated MDFs for both CMDs 5A and 5B are
significantly more metal-poor than this with peaks at
[M/H]$=-1.18^{+0.07}_{-0.04}$ dex and [M/H]$=-1.22^{+0.22}_{-0.04}$
dex, respectively. It therefore seems that the most recent episode of
star formation is being mistaken for a more-metal poor population
owing to the old isochrone grid being used. This is in accordance with
the results presented in Section \ref{sub:ageerr}, where we showed
that for systems younger than $\sim$6 Gyr, the shape of the RGB becomes
heavily dependent on the age, and relatively large errors in the peak
metallicity are expected. 

It is interesting to note that in the case of the smaller error
profile (CMD 5B, Figure \ref{fig:5ab}; lower right), two distinct
peaks in the MDF are recovered through the interpolation. These peaks
correspond roughly to the metallicity of the system during the peaks of the first two
episodes of star formation ([M/H]$\sim$-1.5 dex, 12 Gyr burst; [M/H]$\sim$-1.3
dex, 8 Gyr burst; see Figure \ref{fig:trigauss}). These peaks are well-recovered in
the interpolated MDF, highlighting the high fidelity of the RGB for
metallicity estimates of old stars. On the other hand, the metallicity
of the system during the youngest starburst ([M/H]$\sim$ -1 dex, 4
Gyr; see Figure \ref{fig:trigauss}) completely lacks a peak in the
interpolated MDF, further
illustrating the ineffectiveness of the RGB metallicity estimation
technique for stellar populations younger than 6 Gyr absent of an
accurate age estimate. 

\subsection{The Effects of a Recent Star Formation Episode}
\label{sub:rec}
\begin{figure}
\epsscale{1.2}
\plotone{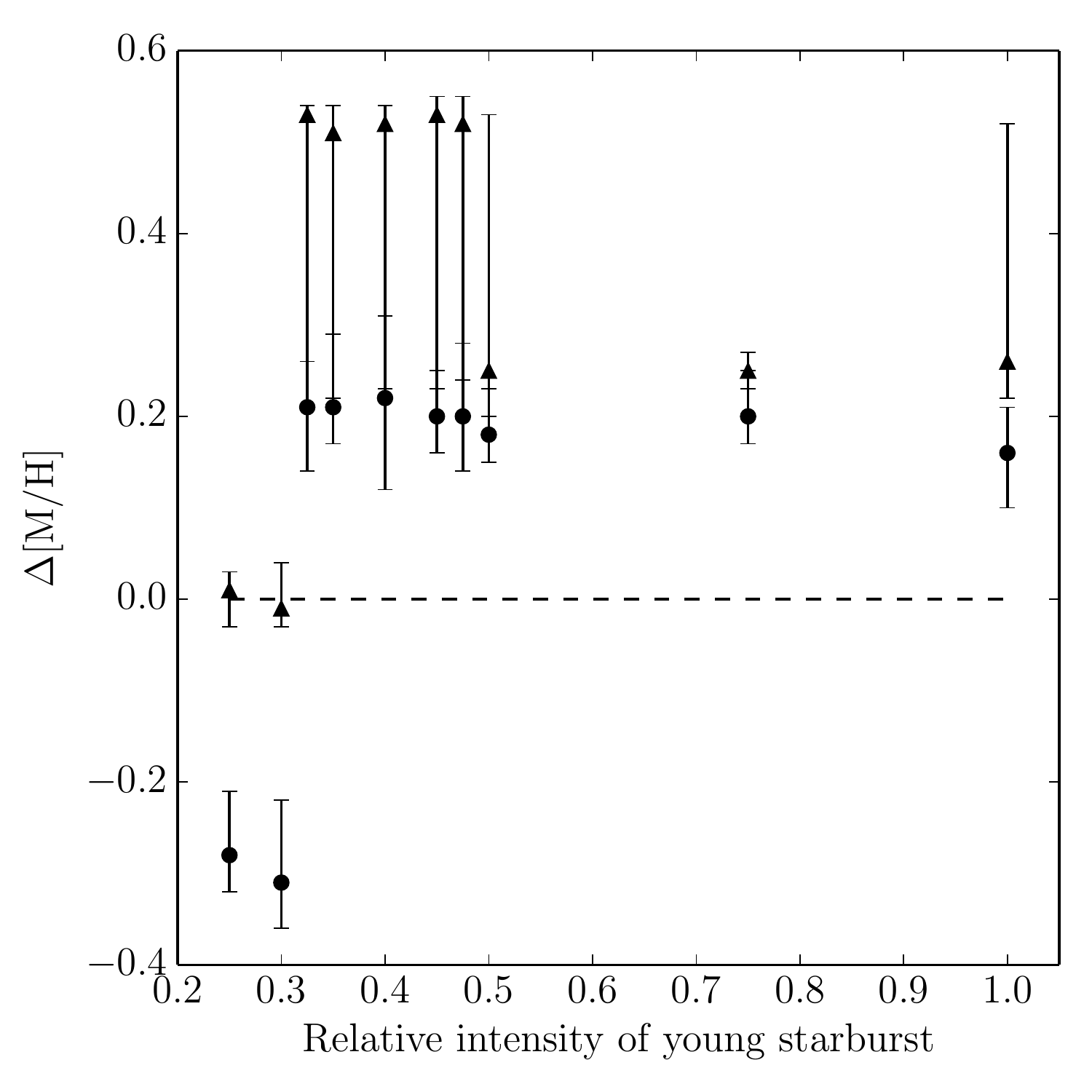}
\caption{\small{Peak metallicity error versus intensity of a young
    starburst (4 Gyr) relative to the old star burst (12 Gyr). CMDs
    produced with observational profile A are circles, while the
    triangles show CMDs produced with observational profile B. Error
    bars represent 99\% confidence intervals returned from
    bootstrapping the peak of the interpolated MDF.\label{fig:young_str_vs_meterr}}}
\end{figure}

We now turn to investigating the effect of a young starburst in more
detail. To accomplish this, we synthesized a series of stellar
populations each with two Gaussian star bursts, one at 12 Gyr,
another at 4 Gyr, and both with 1$\sigma$ widths of 1 Gyr. In each
case of this series, the intensity of the young star formation episode was scaled
down by a multiplicative factor such that its intensity ranged in
0.25-1.0 times the intensity of the old star burst. The AMR for all
populations in this series are identical to P5. The same
procedures for applying the observational profiles and performing the
metallicity interpolation were then performed for each of these CMDs. Figure
\ref{fig:young_str_vs_meterr} shows the error in the peak
metallicity of the interpolated MDF as a function of young starburst
intensity. 

Inspection of Figure \ref{fig:young_str_vs_meterr} reveals that
for young starbursts $\lesssim$0.3 as strong as the old star burst,
the peak of the MDF is well-recovered by interpolating on a grid of
old isochrones. Once this young burst exceeds this threshold
intensity, the young stars begin to dominate the MDF and the peak
interpolated metallicity is in significant error when compared to the
peak of the true MDF. Taking these results with those of the previous
subsection, we conclude that a metallicity determination from the RGB
of a stellar population that has experienced star formation within the
last $\sim$6 Gyr that was at least 30\% as intense as star formation
in older epochs will produce a metallicities which are systematically
in error by up to 0.5 dex.

\subsection{The Effects of Binary Stars}
\label{sub:bin}
Another component of stellar systems that may act to alter an interpolated
MDF is binarity. Studies of the SFH of dwarf galaxies in the Local
Group typically take the binary fraction as a free parameter in their
analysis, and they report values around 40-60\% \citep{hid09,
 mon10, hid11}. We choose to test the affects of binary systems on the
metallicity interpolation by introducing a binary population to
P1. This population contains binary stars in 40\% of the systems with
a minimum mass ratio of $q = 0.5$, with the mass of the secondary
chosen from a flat mass distribution. Subsequent interpolation as with
the other synthetic CMDs did not reveal noticeably different MDFs from
P1, forcing us to conclude that stellar populations with binary
characteristics similar to those used here should not significantly
affect a metallicity interpolation on the RGB.

\section{Conclusions}
\label{sec:con}
We have explored a variety of different stellar population effects on
the MDF constructed from the RGB stars in synthetic CMDs. In
particular, we examine how realistic photometric error and
completeness effects coupled with a variety of SFHs and AMRs alter the
metallicities derived from an interpolation in
M$_{\mathrm I}$, (V-I)$_0$, and $Z$ on the RGB. We summarize our
findings for a variety of simulated model CMDs as follows:
\begin{enumerate}
\item For an old, simple stellar population, the RGB proves to be a
  very reliable metallicity indicator. In the worst case, peak
  metallicity values are recovered within less than 0.1 dex for
  photometry with errors that are relatively large compared with the extent of
  the isochrone grid. In the best case, peak metallicities are
  recovered to within 0.01 dex for more precise photometry. We determine the optimal magnitude range of
  RGB stars over which a metallicity interpolation should be performed
to be restricted to stars between the RGB tip
and  M$_{\mathrm I} = -1.5$ mag for the range of the photometric errors used in this
  exploration. Inherent spreads in the stellar metallicities and ages
  do not have significant effects on the modular interpolated
  metallicity. 
\item It should be noted that the adopted
  isochrone grid can introduce significant systematic effects in the resultant
  MDF. Significant
  selection biases may present themselves in the subsequent
  interpolated MDF resulting from asymmetric scattering of the stars
  off of the model grid.
\item A stellar population significantly younger than 10 Gyr will
  yield systematically more metal-poor metallicities assuming an old
  age for the interpolating isochrones. The resulting
  discrepancy in peak metallicity as a function of age can be
  described with a linear relation. The slope of this relationship
  depends on the photometric error profile of the CMD, indicating that
  more precise photometry is more sensitive to this age error than
  photometry with larger errors. Thus, care should be taken to obtain
  an accurate age estimate when estimating metallicities from the RGB
  using precise photometry, especially if a significant possibility
  exists that the population is younger than 10 Gyr.
\item Using a grid of old isochrones, a star formation episode
  occurring less than $\sim$6 Gyr ago produces a significant,
  systematic error in the interpolated MDF. This can result in
  metallicities that are erroneously poor by up to 0.5 dex. Along the
  same vein, if a recent starburst is $\gtrsim$30\% as intense as older star
  formation events with comparable durations in time, the interpolated
  MDF will yield a significantly erroneous peak.
\end{enumerate}

We close by noting that the RGB of old stellar populations has
proven itself as a reliable metallicity estimator in our study. For
reasonably simple SFHs, the RGB provides a very accurate estimate of
the average metallicity of a stellar population. However, if the
system in question has experienced any significant star formation
within the past few Gyr, metallicities derived from the RGB may be in
error by up to 0.5 dex. Therefore, 
when applied to stellar systems with potentially complex stellar populations, 
more care needs to be
taken if using the RGB to estimate the metallicity of the system. In
particular, a reasonably accurate estimate of the age of the system
should be obtained through an independent, reliable
estimator. Additionally, one should be reasonably certain that no
significant star formation has occurred within the past few Gyr in the
system in question, otherwise there is a high risk of significant systematic errors 
in the derived metallicities.

\acknowledgements

We would like to thank Aaron Grocholski for helping to write the original IDL
interpolation code modified for use in this work, as well as for his
comments on an earlier draft of this paper. We are also grateful to
Mike Barker and Sebastian Hidalgo for their
insightful comments and suggestions on an early draft of this
paper. Finally, we thank the anonymous referee who provided helpful
comments to increase the clarity of the paper. This work has made use of the IAC-STAR Synthetic CMD computation code. IAC-STAR is supported and maintained by the computer division of the Instituto de Astrofísica de Canarias.

\end{document}